\documentclass{Book-Article}     %

\usepackage{graphicx}

\usepackage{amsmath,amsthm}

\newtheorem{theorem}{Theorem}

\theoremstyle{definition}
\newtheorem{definition}{Definition}
\theoremstyle{remark}
\newtheorem{example}{Example}

\usepackage{cite}

\usepackage{hyperref}
\usepackage{xcolor}

\usepackage[irlabel]{fcps}

\usepackage{tempos}

\usepackage{prettyref}
\newcommand{\rref}[2][]{\prettyref{#2}}
\newrefformat{sec}{Section\,\ref{#1}}
\newrefformat{def}{Def.\,\ref{#1}}
\newrefformat{thm}{Theorem\,\ref{#1}}
\newrefformat{prop}{Proposition\,\ref{#1}}
\newrefformat{lem}{Lemma\,\ref{#1}}
\newrefformat{cor}{Corollary\,\ref{#1}}
\newrefformat{ex}{Example\,\ref{#1}}
\newrefformat{tab}{Table\,\ref{#1}}
\newrefformat{fig}{Fig.\,\ref{#1}}
\newrefformat{app}{Appendix\,\ref{#1}}

\providecommand{\bebecomes}{\mathrel{::=}}

\providecommand{\alternative}{~|~}

\newcommand\Ift{\DALint[state=\varphi(t)]}%
\newcommand*{\posterior}[1]{#1^+}

\usepackage{ifpdf}
\ifpdf
\fi

\begin{document}
\begin{frontmatter}          %

\title{Overview of Logical Foundations of Cyber-Physical Systems}
\author{Andr\'e Platzer}
\address{
  Computer Science Department, Carnegie Mellon University, Pittsburgh, USA
  \email{aplatzer@cs.cmu.edu}
}
\runningauthor{A. Platzer}

\begin{abstract}
\emph{Cyber-physical systems} (CPSs) are important whenever computer technology interfaces with the physical world as it does in self-driving cars or aircraft control support systems.
Due to their many subtleties, controllers for cyber-physical systems deserve to be held to the highest correctness standards.
Their correct functioning is crucial, which explains the broad interest in safety analysis technology for their mathematical models, which are called \emph{hybrid systems} because they combine discrete dynamics with continuous dynamics.
\emph{Differential dynamic logic} (\dL) provides logical specification and rigorous reasoning techniques for hybrid systems.
The logic \dL is implemented in the theorem prover \KeYmaeraX, which has been instrumental in verifying ground robot controllers, railway systems, and the next-generation airborne collision avoidance system ACAS~X.
This article provides an informal overview of this logical approach to CPS safety that is detailed in a recent textbook on \emph{Logical Foundations of Cyber-Physical Systems}.
It also explains how safety guarantees obtained in the land of verified models reach the level of CPS execution unharmed.
\end{abstract}



\end{frontmatter}

\section{Introduction}

\emph{Cyber-physical systems} (CPSs) such as aircraft control systems, robots, or partially/fully self-driving cars are very exciting \cite{DBLP:conf/cie/Nerode07,DBLP:conf/hybrid/Pappas11,Alur15}.
They are also formidably challenging to get right due to their physical dynamics and all their subtle interactions with the partially unknown environment.
That is where \emph{differential dynamic logic} \dL \cite{DBLP:journals/jar/Platzer08,Platzer10,DBLP:conf/lics/Platzer12a,DBLP:conf/lics/Platzer12b,DBLP:journals/jar/Platzer17,Platzer18} comes in as a language to express formal specifications for hybrid system models of CPSs.
\emph{Hybrid systems} \cite{DBLP:conf/hybrid/1992} are the basis of all CPS models, because they feature both discrete dynamics (e.g., coming from the stepwise computation and decisions in a computer) and continuous dynamics (e.g., coming from the physical motion of a car).

Differential dynamic logic provides modal formulas \(\dbox{\asprg}{\asfml}\) that express that all behavior of hybrid system $\asprg$ is such that the states $\asprg$ reaches satisfy $\asfml$, which is immediately useful for specification purposes.
For example, the \dL formula \(\asfml\limply\dbox{\asprg}{\bsfml}\) expresses that: if formula $\asfml$ is true initially, then after all runs of hybrid system $\asprg$ is formula $\bsfml$ true.
This particular formula is like a Hoare triple  \(\{\asfml\}\,\asprg\,\{\bsfml\}\) except that it works for hybrid systems instead of conventional discrete programs \cite{DBLP:journals/cacm/Hoare69}.
Yet, differential dynamic logic is more general, because it also makes it possible to state logical formulas of other forms beyond safety.
The \(\dbox{\asprg}{\asfml}\) formula of differential dynamic logic plays a similar role to the \(\tbox{\asfml}\) formula of temporal logic \cite{DBLP:conf/focs/Pnueli77,Prior57} in speaking about $\asfml$ always being true, except that it is parametrized by the particular hybrid systems $\alpha$ after all whose runs $\asfml$ is true.
That makes it possible for \dL formulas to refer to multiple hybrid systems $\asprg,\bsprg$ at once.
The formula \(\dbox{\asprg}{\asfml} \limply \dbox{\bsprg}{\asfml}\), e.g., expresses that if after all runs of $\asprg$ is $\asfml$ true, then also after all runs of $\bsprg$ is $\asfml$ true.
The use of differential dynamic logic is not limited to specification purposes but also provides verification techniques by systematically reducing \dL formulas to simpler \dL formulas by logical transformations until the resulting formulas can be proved.

Fortunately, differential dynamic logic is not just for pen-and-paper proofs but is implemented\footnote{%
The \KeYmaeraX prover is available at \url{http://keymaeraX.org/}
} in the theorem prover \KeYmaeraX \cite{DBLP:conf/cade/FultonMQVP15} that makes it possible to conduct rigorous formal proofs with tool support and substantial levels of proof automation including invariant generation \cite{DBLP:conf/fm/SogokonMTCP19}.
The tool \KeYmaeraX was instrumental for verifying challenging applications, including ground robot obstacle avoidance \cite{DBLP:journals/ijrr/MitschGVP17}, railway systems \cite{DBLP:conf/rssrail/MitschGBGP17}, and the Next-generation Airborne Collision Avoidance System ACAS~X \cite{DBLP:journals/sttt/JeanninGKSGMP17}.
Safety guarantees provided by differential dynamic logic can also transcend models and translate into reality.
\KeYmaeraX implements the transfer of safety proofs for CPS models to CPS implementations by synthesizing provably correct runtime monitors with \emph{ModelPlex} \cite{DBLP:journals/fmsd/MitschP16}, which result in CPS executables that are formally verified in a chain of theorem provers \cite{DBLP:conf/pldi/BohrerTMMP18}.
ModelPlex is also the basis for enabling safe artificial intelligence in cyber-physical systems \cite{DBLP:conf/qest/Platzer19} by wrapping reinforcement learning in a verified safety sandbox \cite{DBLP:conf/aaai/FultonP18} and steering back toward safety when outside the confounds of a well-modeled part of the system behavior \cite{DBLP:conf/tacas/FultonP19}.
These results that will be surveyed here provide rigorous safety guarantees for CPS implementations (even those that involve machine learning) without having to deal with the entire complexity of their implementation during verification.

While a brief overview is given here, significantly more elaborate detail can be found in a recent textbook\footnote{%
Including supporting slides and video lectures are at \url{http://lfcps.org/lfcps/}} on Logical Foundations of Cyber-Physical Systems \cite{Platzer18} as well as a research monograph on Logical Analysis of Hybrid Systems \cite{Platzer10}.
A technical survey of basic differential dynamic logic can be found in a LICS'12 tutorial \cite{DBLP:conf/lics/Platzer12a}.

\section{CPS Programs}

\emph{Hybrid systems} are mathematical models that combine discrete dynamical systems with continuous dynamical systems as is crucial to understand CPSs.
A variety of different types of hybrid system models have been developed, including hybrid automata \cite{DBLP:conf/hybrid/NerodeK92a,DBLP:conf/hybrid/AlurCHH92}, hybrid Petri nets \cite{DBLP:journals/deds/DavidA01}, hybrid CSP \cite{DBLP:conf/hybrid/ChaochenJR95} and hybrid process algebras \cite{DBLP:journals/jlp/CuijpersR05}.
The Logical Foundations of Cyber-Physical Systems are best built on \emph{hybrid programs} \cite{DBLP:journals/jar/Platzer08,Platzer10,DBLP:conf/lics/Platzer12a,DBLP:conf/lics/Platzer12b,DBLP:journals/jar/Platzer17,Platzer18}, which represent hybrid systems in a programming language that includes differential equations.
The relationship of hybrid programs to hybrid automata is similar to that of regular languages to finite automata. The advantage of hybrid programs for verification purposes is that they make it possible to leverage programming language compositionality principles to completely decompose the analysis of hybrid programs into respective analyses of their pieces \cite{DBLP:journals/jar/Platzer08,DBLP:conf/lics/Platzer12b,DBLP:journals/jar/Platzer17}.

\paragraph{Syntax.}
Unlike conventional discrete programs, hybrid programs provide differential equations as primitives for continuous dynamics.
In order to do justice to the demands of cyber-physical systems, hybrid programs feature real number computations and natively offer nondeterminism with which uncertainty about the evolution of the physical world can be expressed.
Such nondeterminism is crucial to accurately capture the fact that we have limited knowledge about what other agents in the physical environment may do (it really is hard to know what exactly the car in front of you will do when).
But nondeterminism also comes in handy when we want to be deliberately imprecise about what happens when in our own control in order to arrive at a more abstract but easier model (even if we could in principle know exactly when our car controller speeds up or slows down as a function of distance, speed limit, gear, gas-air-ratio, engine rotations per minute, and passenger comfort goals, we may be better off retaining a more abstract nondeterministic decision to either accelerate or brake).

\begin{definition}[Hybrid programs] \label{def:HP}
The syntax of \emph{hybrid programs} (HP) is described by the following grammar where $\asprg,\bsprg$ are hybrid programs, $x$ is a variable and $\astrm,\genDE{x}$ are terms, $\ivr$ is a (usually first-order) logical formula:
\begin{equation*}
  \asprg,\bsprg ~\bebecomes~
  \pupdate{\pumod{x}{\astrm}}
  \alternative
  \ptest{\ivr}
  \alternative
  \pevolvein{\D{x}=\genDE{x}}{\ivr}
  \alternative
  \pchoice{\asprg}{\bsprg}
  \alternative
  \asprg;\bsprg
  \alternative
  \prepeat{\asprg}
\end{equation*}
\end{definition}

Polynomial terms $\astrm$ suffice, but rational functions can be used (when not dividing by zero \cite{DBLP:conf/cade/BohrerFP19}).
The \emph{assignment} \(\pupdate{\pumod{x}{\astrm}}\) instantaneously changes the value of variable $x$ to the value of the term $\astrm$ by a discrete jump.
The \emph{test} \(\ptest{\ivr}\) checks if formula $\ivr$ is true in the current state and aborts the execution discarding the run otherwise.
That is, tests have absolutely no effect on the state of the system except that they abort executions that do not pass them.
Such tests are useful to impose conditions on the (especially branching) execution of a system.
Most importantly, the \emph{differential equation} \(\pevolvein{\D{x}=\genDE{x}}{\ivr}\) can be followed for any nondeterministic amount of time (including 0), but only as long as the logical formula $\ivr$ is true at \emph{every} time along the solution of \(\D{x}=\genDE{x}\).
In particular, just like in a failed test, it is impossible to run a differential equation for any amount of time at all unless $\ivr$ is true in the beginning.

The \emph{nondeterministic choice} \(\pchoice{\asprg}{\bsprg}\) can either run the hybrid program $\asprg$ or the program $\bsprg$.
It is the fundamental branching construct, making no prior commitment as to which side is run.
Unlike \(\pchoice{\asprg}{\bsprg}\) which \emph{either} runs $\asprg$ or $\bsprg$,
the \emph{sequential composition} $\asprg;\bsprg$ \emph{first} runs $\asprg$ and then $\bsprg$.
It is the fundamental construct for running things in succession.
Finally, the \emph{nondeterministic repetition} $\prepeat{\asprg}$ will repeatedly run hybrid program $\asprg$ any arbitrary natural number of times, including possibly 0 times.
It is the fundamental construct for repeated actions.

\begin{example}[Bouncing ball] \label{ex:bouncing-ball-HP}
As a simple hybrid system, consider a bouncing ball at altitude $x$ with velocity $v$ that is bouncing up and down over time subject to gravity \(\pevolve{\D{x}=v\syssep\D{v}=-g}\) where $g>0$ is the constant for gravity (\rref{fig:bouncingball-simple-trajectory}).
\begin{figure}[htb]
  \begin{minipage}{5.7cm}
\begin{equation*}
\begin{aligned}
  &\big\lpgroup
  \lpbrace\pevolvein{\D{x}=v\syssep\D{v}=-g}{x\geq0}\rpbrace;\\
  &\quad\lpgroup\pchoice{\ptest{x=0};\pupdate{\pumod{v}{-cv}}}{\ptest{x\neq0}}\rpgroup 
  \prepeat{\big\rpgroup}
\end{aligned}
\end{equation*}
  \end{minipage}
  \qquad
  \begin{minipage}{4cm}
  \includegraphics[height=2.5cm]{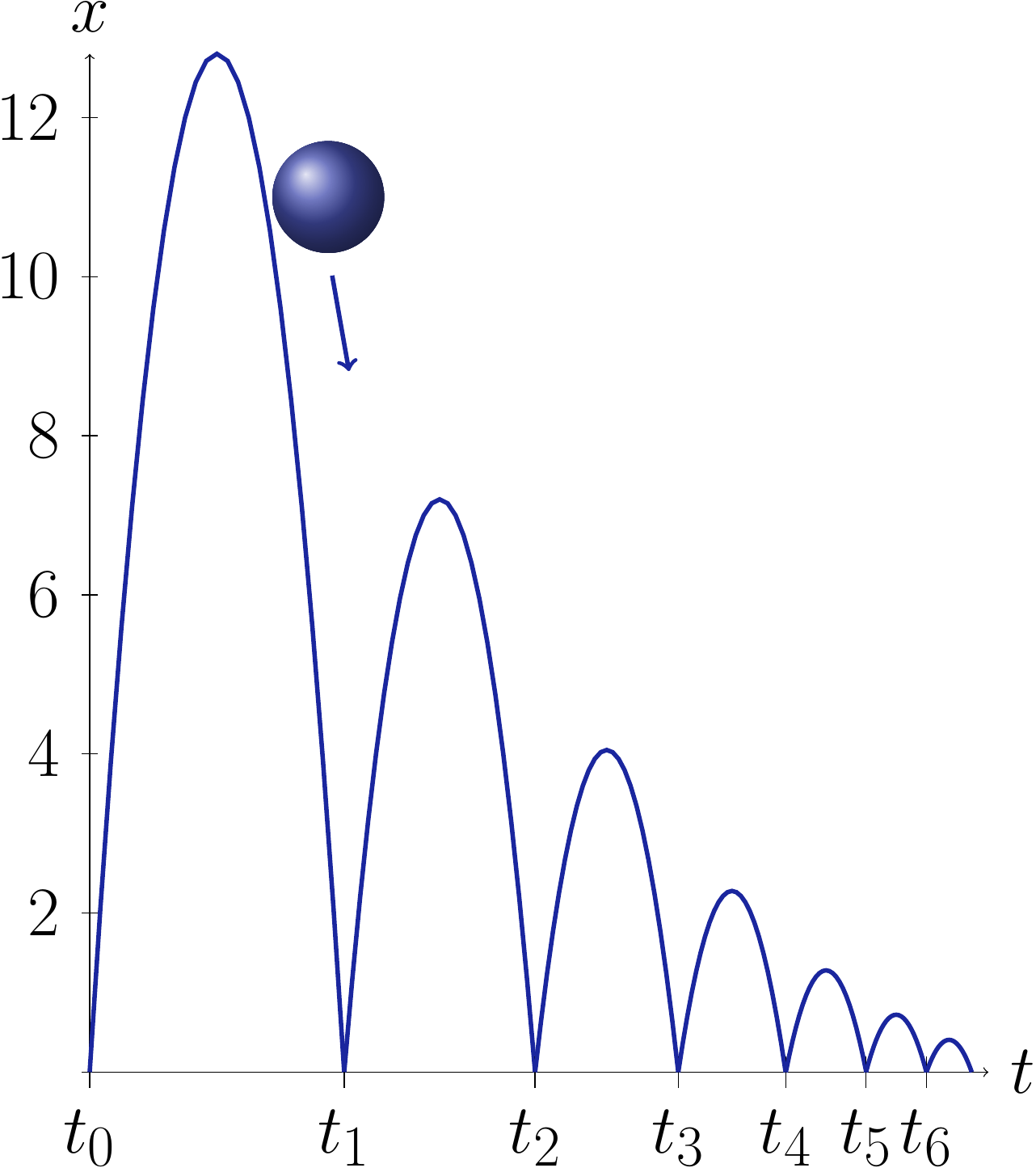}
  \end{minipage}
  \caption{Hybrid program and sample trajectory of a bouncing ball (plotted as height over time)}
  \label{fig:bouncingball-simple-trajectory}
\end{figure}
Since the bouncing ball is unable to fall through the ground at position 0, the differential equation is limited to remain above ground \(\pevolvein{\D{x}=v\syssep\D{v}=-g}{x\geq0}\).
That is, the bouncing ball can continue falling in gravity for any amount of time, but can never fall through the cracks in the ground.
After the bouncing ball stops its continuous evolution, it follows a nondeterministic choice whose right branch tests whether the ball is in the air (with the test \m{\ptest{x\neq0}}) in which case no change happens.
Its left branch tests if the ball lies on the ground (test \m{\ptest{x=0}}) and then inverts the velocity by a discrete assignment (\(\pupdate{\pumod{v}{-cv}}\)) subject to a damping coefficient $0\leq c\leq1$.
That is why the discrete controller in line~2 is can also be written as \(\pifs{x=0}{\pupdate{\pumod{v}{-cv}}}\).
Because the velocity will be flipped from negative to positive when on the ground, that will send the ball flying back up in the air as soon as it next continues along \(\pevolvein{\D{x}=v\syssep\D{v}=-g}{x\geq0}\) after repeating the loop. 
The repetition operator ($\prepeat{}$) at the end of the hybrid program in \rref{fig:bouncingball-simple-trajectory} indicates that it can be repeated any arbitrary number of times.

Note that all choices in all operators of hybrid programs are resolved nondeterministically.
The differential equation can be stopped at any time (just cannot violate the evolution domain constraint \(x\geq0\)).
When stopped above ground (\(x>0\)), the subsequent discrete controller has no effect and the only way forward is to, nondeterministically, stop or repeat the loop and, if it the loop is repeated, continue along the same differential equation in line~1.
At the latest on the ground (\(x=0\)), however, does the differential equation have to stop and give the discrete control a chance to inspect the state and flip the sign of the velocity.
Also note how some choices, e.g., the nondeterministic choice $\pchoice{}{}$ in \rref{fig:bouncingball-simple-trajectory} are nondeterministic but, because of the subsequent tests in both subprograms, really only one choice will succeed in every state.
That is why a nondeterministic default is helpful, because more deterministic behavior can always be defined by imposing suitable tests in the execution of a nondeterministic program. Conversely, a deterministic program cannot properly be made nondeterministic anymore.

While bouncing balls are really no cyber-physical systems (they apparently forgot their computing devices), their behavior can still be explained very well with a prototypical hybrid system where the mix of discrete and continuous dynamics comes from describing the sudden bounce back from the ground versus the continuous flying and falling above ground.
It is much easier to understand bouncing balls from such a hybrid systems perspective than to investigate the fast-paced continuous deformation of the ball on the ground happening at a very different time scale than its continuous flying and falling in gravity.
\end{example}

What is crucial for cyber-physical systems is not only the addition of differential equations, but also to embrace nondeterminism and real-valued quantities.

\paragraph{Semantics.}
The effect of running a hybrid program is that the state of the system changes, which is why the semantics of hybrid programs is defined as the relation it induces between initial and final states.
Because of the nondeterminism, some hybrid programs may reach multiple (or no) final states from a given initial state.
A \emph{state} is a mapping $\iget[state]{\I}$ that assigns a real number $\iget[state]{\I}(x)\in\reals$ to each variable $x$.
The set of all states is denoted $\linterpretations{\Sigma}{\allvars}$.
The real value of a term $\astrm$ in state $\iget[state]{\I}$ is written \(\ivaluation{\I}{\astrm}\) and defined as usual in real arithmetic.

\begin{definition}[Transition semantics]
  \label{def:HP-transition}
  The \emph{semantics} of an HP $\asprg$ defines the transition relation \m{\iaccess[\asprg]{} \subseteq \linterpretations{\Sigma}{\allvars} \times \linterpretations{\Sigma}{\allvars}}
  of initial and final states inductively:
  \[\begin{array}{@{}l@{~}c@{~}l@{}}
    \iaccess[\pupdate{\pumod{x}{\genDJ{x}}}]{\I} &\text{=}& \{(\iget[state]{\I},\iget[state]{\It}) \with \iget[state]{\It}=\iget[state]{\I}~\text{except that}~ \ivaluation{\It}{x}=\ivaluation{\I}{\genDJ{x}}\}\\
  \iaccess[\ptest{\ivr}]{\I} &\text{=}& \{(\iget[state]{\I},\iget[state]{\I}) \with \imodels{\I}{\ivr}\}\\
  \iaccess[\pevolvein{\D{x}=\genDE{x}}{\ivr}]{\I} &\text{=}& \big\{(\iget[state]{\I},\iget[state]{\It}) :
        \iget[flow]{\If}(0)=\iget[state]{\I} ~\text{except at $\D{x}$ and}~ \iget[flow]{\If}(r)=\iget[state]{\It}
        ~\text{for a solution}\\
        &&\text{\m{\iget[flow]{\If}{:}[0,r]\to\linterpretations{\Sigma}{\allvars}} of any duration $r$ satisfying}~
        \imodels{\If}{\D{x}=\genDE{x} \land \ivr}
        \big\}\\
  \iaccess[\pchoice{\asprg}{\bsprg}]{\I} &\text{=}& \iaccess[\asprg]{\I} \cup \iaccess[\bsprg]{\I}\\
  \iaccess[\asprg;\bsprg]{\I} &\text{=}& \iaccess[\asprg]{\I} \compose\iaccess[\bsprg]{\I}
= \{(\iget[state]{\I},\iget[state]{\It}) : (\iget[state]{\I},\iget[state]{\Iz}) \in \iaccess[\asprg]{\I},  (\iget[state]{\Iz},\iget[state]{\It}) \in \iaccess[\bsprg]{\I}\}\\
  \iaccess[\prepeat{\asprg}]{\I} &\text{=}& \closureTransitive{\iaccess[\asprg]{\I}} = \displaystyle\cupfold_{n\in\naturals}\iaccess[{\prepeat[n]{\asprg}}]{\I}
~\text{with}~\prepeat[n+1]{\asprg} \mequiv \prepeat[n]{\asprg};\asprg ~\text{and}~ \prepeat[0]{\asprg}\mequiv\,\ptest{\ltrue}
  \end{array}\]
where \m{\imodels{\If}{\D{x}=\genDE{x} \land \ivr}},
iff for all times \m{0\leq\intime\leq r}: \m{\imodels{\Iff[\intime]}{\D{x}=\genDE{x} \land \ivr}} 
with \m{\iget[state]{\Iff[\intime]}(\D{x}) \mdefeq \D[t]{\iget[state]{\Ift}(x)}(\intime)}
and \(\iget[state]{\Iff[\intime]}=\iget[state]{\Iff[0]}\) except at $x,\D{x}$.
\end{definition}

That is, the relation \m{\imodels{\If}{\D{x}=\genDE{x} \land \ivr}} holds when $\iget[flow]{\If}$ is a solution of the differential equation \(\D{x}=\genDE{x}\) that respects the evolution domain constraint $\ivr$ at every point in time and, of course, no other variable changes except those indicated by the differential equation.
The semantics is explicit change: nothing changes unless the hybrid program indicates how.

The semantics \(\iaccess[\pchoice{\asprg}{\bsprg}]{\I}\) of a nondeterministic choice \(\pchoice{\asprg}{\bsprg}\) is the union \(\iaccess[\asprg]{\I} \cup \iaccess[\bsprg]{\I}\) of the individual semantics', which is where the notation comes from.
The semantics \(\iaccess[\asprg;\bsprg]{\I}\) of a sequential composition \(\asprg;\bsprg\) is the relation composition \(\iaccess[\asprg]{\I} \compose\iaccess[\bsprg]{\I}\).
The semantics \(\iaccess[\prepeat{\asprg}]{\I}\) of a nondeterministic repetition is the reflexive transitive closure \(\closureTransitive{\iaccess[\asprg]{\I}}\) of the semantics \(\iaccess[\asprg]{\I}\) of $\asprg$ or, equivalently, the semantics of all the $n$-fold iterations $\prepeat[n]{\asprg}$ of $\asprg$.
The semantics of an assignment \(\pupdate{\pumod{x}{\genDJ{x}}}\) changes the value of the variable $x$ in the new state $\iget[state]{\It}$ to the real value \(\ivaluation{\I}{\genDJ{x}}\) that the right-hand side term $\genDJ{x}$ had in the old state $\iget[state]{\I}$.
The semantics \(\iaccess[\ptest{\ivr}]{\I}\) of a test $\ptest{\ivr}$ does not change the state but a transition is only possible in a state $\iget[state]{\I}$ that is in the set of all states in which $\ivr$ is true (written $\imodel{\I}{\ivr}$ and defined in \rref{def:dL-semantics} below).

\section{Differential Dynamic Logic}

\emph{Differential dynamic logic} \dL \cite{DBLP:journals/jar/Platzer08,Platzer10,DBLP:conf/lics/Platzer12a,DBLP:conf/lics/Platzer12b,DBLP:journals/jar/Platzer17,Platzer18} provides specification and verification techniques for hybrid systems written as hybrid programs.

\paragraph{Syntax.}
The defining constructs of differential dynamic logic are the pair of modalities $\dbox{\asprg}{}$ and $\ddiamond{\asprg}{}$ for every hybrid program $\asprg$, which can be used in front of any \dL formula $\asfml$.
The \dL formula \(\dbox{\asprg}{\asfml}\) is true in any state in which $\asfml$ is true after all runs of hybrid program $\asprg$.
The \dL formula \(\ddiamond{\asprg}{\asfml}\) is true if there is a run of hybrid program $\asprg$ after which $\asfml$ is true.
The rest of \dL is built like first-order logic of real arithmetic including quantifiers over the reals.
Real arithmetic makes the most sense when talking about the real positions and velocities of CPSs.

\begin{definition}[\dL formulas] \label{def:dL-formula}
  The syntax of \emph{formulas of differential dynamic logic} is described by the following grammar
  where $\asfml,\bsfml$ are formulas, $\astrm,\bstrm$ are terms, $x$ is a variable and $\asprg$ is a hybrid program:
\begin{equation*}
  \asfml,\bsfml ~\bebecomes~
  \astrm=\bstrm \alternative
  \astrm\geq\bstrm \alternative
  \lnot \asfml \alternative
  \asfml \land \bsfml \alternative
  \asfml \lor \bsfml \alternative
  \asfml \limply \bsfml \alternative
  \lforall{x}{\asfml} \alternative 
  \lexists{x}{\asfml} \alternative
  \dbox{\asprg}{\asfml}
  \alternative \ddiamond{\asprg}{\asfml}
\end{equation*}
\end{definition}
Bi-implications \(\asfml\lbisubjunct\bsfml\) are defined syntactically as mutual implication \((\asfml\limply\bsfml)\land(\bsfml\limply\asfml)\).
Likewise, other operators such as $>,\leq,<,\neq$ are definable as usual.

By combining the operators it is easy to express various correctness statements about hybrid systems.
For example, a \dL formula of the shape \(\asfml \limply \dbox{\asprg}{\bsfml}\)
says that if $\asfml$ is true then all runs of $\asprg$ satisfy $\bsfml$.
Hence, \(\asfml \limply \dbox{\asprg}{\bsfml}\) is valid (true in all states) when formula $\bsfml$ is, indeed, true after all runs of hybrid system $\asprg$ that start in an initial state where formula $\asfml$ is true (safety).
Likewise, \(\asfml \limply \ddiamond{\asprg}{\bsfml}\)
says that if $\asfml$ is true then there is a run of $\asprg$ after which $\bsfml$ is true (liveness).
Other correctness properties mix multiple modalities in the same formula.
For example, \(\asfml \limply \dbox{\asprg}{\ddiamond{\bsprg}{\bsfml}}\)
expresses that if $\asfml$ is true (initially), then after all behaviors of $\asprg$, there is a run of $\bsprg$ that makes $\bsfml$ true in the end.
The formula \(\asfml \limply (\dbox{\asprg}{\bsfml} \lbisubjunct \dbox{\bsprg}{\bsfml})\) expresses a conditional equivalence that if $\asfml$ is true (initially), then all behaviors of $\asprg$ satisfy $\bsfml$ iff all behaviors of $\bsprg$ satisfy $\bsfml$.
Quantifiers can be useful, e.g., to say that for all (initial) positions $x$ there is an (initial) velocity $v$ for which all behavior of car $\asprg$ obeys a safety property $\asfml$ as \(\lforall{x}{\lexists{v}{\dbox{\asprg}{\asfml}}}\).

\begin{example}[Bouncing ball conjecture] \label{ex:bouncing-ball-conjecture}
Continuing with the hybrid program from \rref{ex:bouncing-ball-HP}, a \dL formula expressing safety of the bouncing ball is the following:
\begin{multline*}
0\leq x \land x=H \limply\\
  \big[\big\lpgroup
  \lpbrace\pevolvein{\D{x}=v\syssep\D{v}=-g}{x\geq0}\rpbrace;
  ~(\pchoice{\ptest{x=0};\pupdate{\pumod{v}{-cv}}}{\ptest{x\neq0}}) 
  \prepeat{\big\rpgroup}
  \big]\,(0\leq x \land x \leq H)
\end{multline*}
It expresses that if the bouncing ball starts at the initial height $H$ above ground, then it will always remain above ground but below initial height (\(0\leq x\land x\leq H\)).
Unfortunately, no matter how intuitive such a claim may look like at first glance, the above \dL formula is not actually true for all (initial) values of $x,v,g,c,H$.
Of course, if the initial values of $x$ and $H$ disagree, then the formula is vacuously true, because the left-hand side of its implication is false.
But additional assumptions are also needed on the other variables as well to make the formula true for all (initial) values of all variables.
Which assumptions are those?
\end{example}

\paragraph{Semantics.}
Strictly speaking, the definitions of hybrid programs (\rref{def:HP}) and of differential dynamic logic formulas (\rref{def:dL-formula}) are mutually inductive, because hybrid programs occur in modal formulas while formulas occur in tests or evolution domain constraints.
That is why their semantics is also defined by mutual induction (\rref{def:HP-transition} and \rref{def:dL-semantics}).
But since not much intuition is lost when first imagining only simple arithmetic formulas can occur in hybrid programs (so-called poor test \dL), this subtlety is often glossed over in the interest of a simple presentation.

The meaning of a \dL formula is the set of all states in which it is true (from which the set of all states in which it is false can be read off by complementation).

\begin{definition}[Semantics]
  \label{def:dL-semantics}
  The \emph{semantics of formulas} is defined inductively, for each \dL formula $\asfml$, as the set of states, written \m{\imodel{\I}{\asfml} \subseteq \linterpretations{\Sigma}{\allvars}}, in which $\asfml$ is true:
  \[\begin{array}{l@{~}c@{~}l}
    \imodel{\I}{\astrm=\bstrm} &\text{=}& \{\iget[state]{\I} \with \ivaluation{\I}{\astrm} = \ivaluation{\I}{\bstrm}\}\\
    \imodel{\I}{\astrm\geq\bstrm} &\text{=}& \{\iget[state]{\I} \with \ivaluation{\I}{\astrm} \geq \ivaluation{\I}{\bstrm}\}\\
    \imodel{\I}{\lnot\asfml} &\text{=}& \scomplement{\imodel{\I}{\asfml}} = \linterpretations{\Sigma}{\allvars} \setminus \imodel{\I}{\asfml}\\
    \imodel{\I}{\asfml\land\bsfml} &\text{=}& \imodel{\I}{\asfml} \cap \imodel{\I}{\bsfml}\\
    \imodel{\I}{\asfml\lor\bsfml} &\text{=}& \imodel{\I}{\asfml} \cup \imodel{\I}{\bsfml}\\
    \imodel{\I}{\asfml\limply\bsfml} &\text{=}& \scomplement{\imodel{\I}{\asfml}} \cup \imodel{\I}{\bsfml}\\
    \imodel{\I}{\lexists{x}{\asfml}} &\text{=}& \{\iget[state]{\I} \,:\, \imodels{\It}{\asfml} ~\text{for some state $\iget[state]{\It}$ that agrees with $\iget[state]{\I}$ except on $x$}\}\\
    \imodel{\I}{\lforall{x}{\asfml}} &\text{=}& \{\iget[state]{\I} \,:\, \imodels{\It}{\asfml} ~\text{for all states $\iget[state]{\It}$ that agree with $\iget[state]{\I}$ except on $x$}\}\\
    \imodel{\I}{\ddiamond{\asprg}{\asfml}} &\text{=}& %
    \{\iget[state]{\I} \,:\, \imodels{\It}{\asfml} ~\text{for some state $\iget[state]{\It}$ such that}~ \iaccessible[\asprg]{\I}{\It}\}\\
    \imodel{\I}{\dbox{\asprg}{\asfml}} &\text{=}& %
    \{\iget[state]{\I} \,:\, \imodels{\It}{\asfml} ~\text{for all states $\iget[state]{\It}$ such that}~ \iaccessible[\asprg]{\I}{\It}\}
  \end{array}\]
  A \dL formula $\asfml$ is \emph{valid}, written \(\entails\asfml\), iff it is true in all states, i.e., \(\imodel{\I}{\asfml}=\linterpretations{\Sigma}{\allvars}\).
\end{definition}

The formula \(\dbox{\asprg}{\asfml}\) is true in any state $\iget[state]{\I}$ from which all states $\iget[state]{\It}$ reachable after running hybrid program $\asprg$ satisfy $\asfml$.
The formula \(\ddiamond{\asprg}{\asfml}\) is true in any state $\iget[state]{\I}$ from which there is a state $\iget[state]{\It}$ that is reachable after running hybrid program $\asprg$ and that satisfies $\asfml$.

Valid formulas are the most valuable ones, because we can rely on them being true no matter what state a system is presently in.
For example, an implication \(\asfml\limply\bsfml\) is valid iff \(\imodel{\I}{\asfml}\subseteq\imodel{\I}{\bsfml}\), i.e., the set of all states in which $\asfml$ is true is a subset of the set of all states in which $\bsfml$ is true.
A bi-implication \(\asfml\lbisubjunct\bsfml\) is valid iff \(\imodel{\I}{\asfml}=\imodel{\I}{\bsfml}\), i.e., the set of all states in which $\asfml$ is true is the same as the set of all states in which $\bsfml$ is true.

\begin{example}[Bouncing ball safety] \label{ex:bouncing-ball-dL}
Augmenting \rref{ex:bouncing-ball-conjecture} with sufficiently many assumptions on other variables to make the formula about the behavior of bouncing balls valid (so true in all states) leads, e.g., to the following \dL formula:
\begin{multline*}
0\leq x \land x=H \land v=0 \land g>0 \land 1=c \limply\\
  \big[\big\lpgroup
  \lpbrace\pevolvein{\D{x}=v\syssep\D{v}=-g}{x\geq0}\rpbrace;
  ~(\pchoice{\ptest{x=0};\pupdate{\pumod{v}{-cv}}}{\ptest{x\neq0}}) 
  \prepeat{\big\rpgroup}
  \big]\,(0\leq x \land x \leq H)
\end{multline*}
Of course, slightly more general \dL formulas are valid as well.
But assuming damping coefficient $c=1$, for example, has the pleasant effect of considerably simplifying notation subsequently.
\end{example}

\section{Dynamic Axioms for Dynamical Systems}

Definitions~\ref{def:HP-transition} and~\ref{def:dL-semantics} give \dL a rigorous mathematical semantics but, except in simple cases, working through their uncountable sets of states does not exactly give a particularly practical way of establishing whether formulas are true in a state let alone valid (so true in all states).
Differential dynamic logic is not just useful for specifying CPS but also for verifying CPS thanks to its proof calculus, with which the validity of formulas can be established by a finite syntactic proof instead of the uncountably infinite sets of states in the semantics.

Axioms and proof rules for differential dynamic logic \cite{DBLP:conf/lics/Platzer12b,DBLP:journals/jar/Platzer17} are listed in \rref{fig:dL}.
The usual axioms and proof rules for propositional connectives and first-order logic can be used as a basis \cite{DBLP:journals/jar/Platzer08,Platzer18} but are not discussed further here.

\begin{figure}[tbhp]
  \centering
  \begin{calculus}
\cinferenceRule[diamond|$\didia{\cdot}$]{diamond axiom}
{\linferenceRule[equiv]
  {\lnot\dbox{\ausprg}{\lnot \ausfml}}
  {\axkey{\ddiamond{\ausprg}{\ausfml}}}
}
{}
\cinferenceRule[assignb|$\dibox{:=}$]{assignment / substitution axiom}
{\linferenceRule[equiv]
  {p(\genDJ{x})}
  {\axkey{\dbox{\pupdate{\umod{x}{\genDJ{x}}}}{p(x)}}}
}
{}%
\cinferenceRule[testb|$\dibox{?}$]{test}
{\linferenceRule[equiv]
  {(\ivr \limply \ausfml)}
  {\axkey{\dbox{\ptest{\ivr}}{\ausfml}}}
}{}%
\cinferenceRule[evolveb|$\dibox{'}$]{evolve}
{\linferenceRule[equiv]
  {\lforall{t{\geq}0}{\dbox{\pupdate{\pumod{x}{\solf(t)}}}{p(x)}}}
  {\axkey{\dbox{\pevolve{\D{x}=\genDE{x}}}{p(x)}}}
  \hspace{0.75cm}
}{\m{\D{\solf}(t)=\genDE{\solf}}}%
\cinferenceRule[choiceb|$\dibox{\cup}$]{axiom of nondeterministic choice}
{\linferenceRule[equiv]
  {\dbox{\ausprg}{\ausfml} \land \dbox{\busprg}{\ausfml}}
  {\axkey{\dbox{\pchoice{\ausprg}{\busprg}}{\ausfml}}}
}{}%
\cinferenceRule[composeb|$\dibox{{;}}$]{composition} %
{\linferenceRule[equiv]
  {\dbox{\ausprg}{\dbox{\busprg}{\ausfml}}}
  {\axkey{\dbox{\ausprg;\busprg}{\ausfml}}}
}{}%
\cinferenceRule[iterateb|$\dibox{{}^*}$]{iteration/repeat unwind} %
{\linferenceRule[equiv]
  {\ausfml \land \dbox{\ausprg}{\dbox{\prepeat{\ausprg}}{\ausfml}}}
  {\axkey{\dbox{\prepeat{\ausprg}}{\ausfml}}}
}{}%
\cinferenceRule[K|K]{K axiom / modal modus ponens}
{\linferenceRule[impl]
  {\dbox{\ausprg}{(\ausfml\limply\busfml)}}
  {(\dbox{\ausprg}{\ausfml}\limply\axkey{\dbox{\ausprg}{\busfml}})}
}{}%
\cinferenceRule[Ieq|I]{loop induction}
{\linferenceRule[equiv]
  {\ausfml \land \dbox{\prepeat{\ausprg}}{(\ausfml\limply\dbox{\ausprg}{\ausfml})}}
  {\axkey{\dbox{\prepeat{\ausprg}}{\ausfml}}}
}{}%
\cinferenceRule[V|V]{vacuous $\dbox{}{}$}
{\linferenceRule[impl]
  {p}
  {\axkey{\dbox{\ausprg}{p}}}
}{\m{FV(p)\cap BV(\ausprg)=\emptyset}}%
\cinferenceRule[G|G]{$\dbox{}{}$ generalization} %
{\linferenceRule[formula]
  {\ausfml}
  {\dbox{\ausprg}{\ausfml}}
}{}%

  \end{calculus}
  \caption{Differential dynamic logic axioms and proof rules}
  \label{fig:dL}
\end{figure}

\subsection{Axioms}
The idea behind the axioms in \rref{fig:dL} is that, when used to (equivalently) reduce the (usually left) \axkey{blue} side to the rest, they symbolically decompose a property of a larger hybrid program into a range of properties of its pieces.
For example, the axiom of nondeterministic choice \irref{choiceb} is an equivalence that makes it possible to replace any formula of the shape \(\dbox{\pchoice{\ausprg}{\busprg}}{\ausfml}\) equivalently with a corresponding conjunction \(\dbox{\ausprg}{\ausfml} \land \dbox{\busprg}{\ausfml}\) that expresses safety of smaller subprograms.
Since both sides are equivalent according to axiom \irref{choiceb}, replacing one side by the other in any context does not change the overall truth.
But the right-hand side of axiom \irref{choiceb} consists of structurally simpler formulas, because all its hybrid programs are simpler than those on the left-hand side, so that using axiom \irref{choiceb} to replace the left by the right side makes progress in simplifying formulas.
Depending on the polarity (number of negations in negation normal form) with which \(\dbox{\pchoice{\ausprg}{\busprg}}{\ausfml}\) occurs in a formula, either the right-to-left or left-to-right implication of axiom \irref{choiceb} is needed to justify correctness of that transformation.
Since axiom \irref{choiceb}, like most others, is an equivalence axiom, it is easy to see (and easy to prove \cite{DBLP:journals/jar/Platzer17}) why any occurrence of a formula of the form \(\dbox{\pchoice{\ausprg}{\busprg}}{\ausfml}\) can be replaced equivalently by the corresponding equivalent \(\dbox{\ausprg}{\ausfml} \land \dbox{\busprg}{\ausfml}\) regardless of the context.

The sequential composition axiom \irref{composeb} performs a more subtle simplifying decomposition by making it possible to equivalently replace formulas of the form \(\dbox{\ausprg;\busprg}{\ausfml}\) with more complex hybrid programs into nested modalities \(\dbox{\ausprg}{\dbox{\busprg}{\ausfml}}\) with more complex postconditions but smaller hybrid programs (no $;$ anymore).

The assignment axiom \irref{assignb} has no more subprograms to work with but expresses that \(p(x)\) is true after assigning \(\pupdate{\pumod{x}{\genDJ{x}}}\) iff \(p(\genDJ{x})\) is true before.
Tests are handled by axiom \irref{testb} which expresses that $\ausfml$ holds after running test $\ptest{\ivr}$ iff \(\ivr\limply\ausfml\) holds now, because the test can only run successfully when $\ivr$ was indeed true.
Axiom \irref{evolveb} replaces differential equations equivalently by a universally quantified property of their solution $\solf$ satisfying \(\D{\solf}(t)=\genDE{\solf}\) and $\solf(0)=x$.
This works for simple differential equations whose solutions are expressible in real arithmetic (and continues to work with slight modifications in the presence of evolution domain constraints), but is hopeless for ``serious'' differential equations.

More involved differential equations use inductive proof techniques based on \emph{differential invariants} \cite{DBLP:journals/logcom/Platzer10,Platzer10,DBLP:journals/jar/Platzer17} that can prove all their real arithmetic invariants in differential dynamic logic \cite{DBLP:conf/lics/PlatzerT18}.
The basic idea of differential invariants is to prove them inductively by proving that their derivatives along the differential equation satisfy the same logical relations.
Other steps in such proofs successively accumulate knowledge about the evolution of the differential equation in its evolution domain constraint (\emph{differential cut}) or soundly edit the differential equation (\emph{differential ghost}) as elaborated elsewhere \cite{DBLP:conf/lics/Platzer12a,DBLP:journals/jar/Platzer17,DBLP:conf/lics/PlatzerT18}.

\subsection{Proof Rules}

Loops can be unwound finitely with the iteration axiom \irref{iterateb} or proved by the induction axiom \irref{Ieq}.
As usual, the postcondition $\ausfml$ may need to be strengthened to identify a proper loop invariant $\inv$.
This strengthening ultimately uses Kripke's modal modus ponens axiom \irref{K} or G\"odel's generalization rule \irref{G}, but is also packaged up with a derived proof rule in sequent calculus:
\[
\dinferenceRule[loop|loop]{inductive invariant}
{\linferenceRule[sequent]
  {\lsequent[L]{} {\inv}
  &\lsequent[g]{\inv} {\dbox{\ausprg}{\inv}}
  &\lsequent[g]{\inv} {\ausfml}}
  {\lsequent[L]{} {\dbox{\prepeat{\ausprg}}{\ausfml}}}
}{}%
\]
As usual in classical logics, the meaning of a \emph{sequent} \(\lsequent{\Gamma}{\Delta}\) is the meaning of the formula \((\landfold_{\asfml\in\Gamma} \asfml) \limply (\lorfold_{\bsfml\in\Delta} \bsfml)\).
That is, the meaning of a comma in the list of assumptions or \emph{antecedent} $\Gamma$ is conjunctive while the meaning of a comma in the \emph{succedent} $\Delta$ is disjunctive.
Sequents are particularly useful for normalizing the shape of logical formulas.

Finally, axiom \irref{diamond} uses dualities of modalities to make duals of all reasoning principles for box modalities from the other axioms available for diamond modalities.
The vacuous axiom \irref{V} is special in that it is only applicable if the free variables of $p$ do not intersect the bound variables written by $\ausprg$, i.e., \(\freevars{p}\cap\boundvars{\ausprg}=\emptyset\).
It shows in a single step that the truth of formulas $p$ remains unaffected by running $\ausprg$ if $p$ does not depend on any variable that program $\ausprg$ changes.
A more refined \emph{uniform substitution} calculus for \dL \cite{DBLP:journals/jar/Platzer17} completely avoids any such side conditions in the entire calculus.
Uniform substitutions make logics significantly easier to implement and improve robustness, which is why the implementation of \dL is entirely based on uniform substitutions \cite{DBLP:conf/cade/FultonMQVP15}.

The differential dynamic logic proof calculus successively reduces properties of hybrid systems to formulas of real arithmetic, which is a surprisingly tame logic where validity is decidable \cite{tarski_decisionalgebra51,DBLP:conf/automata/Collins75,DBLP:journals/jsc/CollinsH91}, although in general of doubly exponential complexity in the number of quantifier block alternations \cite{DBLP:journals/jsc/Weispfenning88,DBLP:journals/jsc/DavenportH88}.

\begin{example}[Bouncing ball proof] \label{ex:bouncing-ball-dL-proof}
\newcommand*{\prehv}[2]{A}%
\newcommand*{\posthv}[2]{B{\scriptstyle(#1,#2)}}%
\newcommand*{\prehvdef}[2]{0\leq #1 \land #1=H \land #2=0 \land g>0 \land 1=c}%
\newcommand*{\posthvdef}[2]{0\leq #1 \land #1 \leq H}%
\newcommand*{\odehv}[2]{\D[2]{#1}..}%
\newcommand*{\odehvdef}[2]{\pevolvein{\D{#1}=#2\syssep\D{#2}=-g}{#1\geq0}}%
\newcommand*{\invhv}[2]{\text{j}{\scriptstyle(#1,#2)}}%
\newcommand*{\invhvdef}[2]{2g#1=2gH-{#2}^2\land #1\geq0}%
The conjectured validity of the bouncing ball \dL formula from \rref{ex:bouncing-ball-dL}, of course, deserves a proper \dL proof as justification.
It is easy to conduct such a proof after choosing a suitable loop invariant $\invhv{x}{v}$.
The proof uses abbreviations for the initial condition ($\prehv{x}{v}$), postcondition ($\posthv{x}{v}$), and for the differential equation with its evolution domain constraint ($\odehv{x}{v}$):
\begin{align*}
  \prehv{x}{v} &\mdefequiv \prehvdef{x}{v}\\
  \posthv{x}{v} &\mdefequiv \posthvdef{x}{v}\\
  \odehv{x}{v} &\mdefequiv \lpbrace\odehvdef{x}{v}\rpbrace
\end{align*}
The first step of the proof (starting at the desired conclusion at the very bottom) is to use a propositional implication rule \irref{implyr} to assume the left-hand side of the implication ($\prehv{x}{v}$) in the antecedent and proceed to prove its right-hand side.
Using rule \irref{loop} with a loop invariant $\invhv{x}{v}$ (identified later), postcondition $\posthv{x}{v}$ of the nondeterministic repetition is proved by showing that $\invhv{x}{v}$ follows from the initial condition $\prehv{x}{v}$ (left premise), inductively remains true when running a round of the loop (middle premise), and finally implies the  postcondition (right premise):
\begin{sequentdeduction}[array]
\linfer[loop]
{\lsequent{\prehv{x}{v}} {\invhv{x}{v}}
!\lsequent{\invhv{x}{v}} {\dbox{\pevolve{\odehv{x}{v}}; (\pchoice{\ptest{x=0};\pupdate{\pumod{v}{-cv}}}{\ptest{x\neq0}})}{\invhv{x}{v}}}
!\lsequent{\invhv{x}{v}} {\posthv{x}{v}}
}
{\linfer[implyr]
{\lsequent{\prehv{x}{v}} {
\dbox{\prepeat{\big(\pevolve{\odehv{x}{v}}; (\pchoice{\ptest{x=0};\pupdate{\pumod{v}{-cv}}}{\ptest{x\neq0}})\big)}}{\posthv{x}{v}}}}
{\lsequent{} {\prehv{x}{v} \limply
\dbox{\prepeat{\big(\pevolve{\odehv{x}{v}}; (\pchoice{\ptest{x=0};\pupdate{\pumod{v}{-cv}}}{\ptest{x\neq0}})\big)}}{\posthv{x}{v}}}}
}
\end{sequentdeduction}
\begin{figure}[htbp]
\renewcommand{\linferPremissSeparation}{~~~~}%
\begin{sequentdeduction}[default]
\linfer[composeb]
 {\linfer[Mbr]
  {\lsequent{\invhv{x}{v}} {\dbox{\pevolve{\odehv{x}{v}}}{\invhv{x}{v}}
  ~}
  !\linfer[choiceb]
    {\linfer[andr]
      {\linfer[composeb]
        {\linfer[testb+implyr]
          {\linfer[assignb]
          {\lsequent{\invhv{x}{v},x=0} {\invhv{x}{-cv}}}
          {\lsequent{\invhv{x}{v},x{=}0} {\dbox{\pupdate{\pumod{v}{-cv}}}{\invhv{x}{v}}}}
          }
        {\lsequent{\invhv{x}{v}} {\dbox{\ptest{x{=}0}}{\dbox{\pupdate{\pumod{v}{-cv}}}{\invhv{x}{v}}}}}
        }
        {\lsequent{\invhv{x}{v}} {\dbox{\ptest{x{=}0};\pupdate{\pumod{v}{-cv}}}{\invhv{x}{v}}}}
      ! \linfer[testb]
      {\lsequent{\invhv{x}{v},x{\neq}0} {\invhv{x}{v}}}
      {\lsequent{\invhv{x}{v}} {\dbox{\ptest{x{\neq}0}}{\invhv{x}{v}}}}
      }
      {\lsequent{\invhv{x}{v}} {\dbox{\ptest{x=0};\pupdate{\pumod{v}{-cv}}}{\invhv{x}{v}}\land\dbox{\ptest{x\neq0}}{\invhv{x}{v}}}}
    }
    {\lsequent{\invhv{x}{v}} {\dbox{\pchoice{\ptest{x=0};\pupdate{\pumod{v}{-cv}}}{\ptest{x\neq0}}}{\invhv{x}{v}}}}
    }
    {\lsequent{\invhv{x}{v}} {\dbox{\pevolve{\odehv{x}{v}}}{\dbox{\pchoice{\ptest{x=0};\pupdate{\pumod{v}{-cv}}}{\ptest{x\neq0}}}{\invhv{x}{v}}}}}
  }
  {\lsequent{\invhv{x}{v}} {\dbox{\pevolve{\odehv{x}{v}}; (\pchoice{\ptest{x=0};\pupdate{\pumod{v}{-cv}}}{\ptest{x\neq0}})}{\invhv{x}{v}}}}
\end{sequentdeduction}
  \caption{Sequent calculus proof shape for induction step of bouncing ball}
  \label{fig:dL-bouncing-ball-proof}
\end{figure}%
The proof of the middle premise continues in \rref{fig:dL-bouncing-ball-proof}.
That proof is follow-your-nose except that it saves some writing by using a derived monotonicity rule that makes it possible to replace postcondition $\ausfml$ by a postcondition $\busfml$ implying $\ausfml$:
\[
\dinferenceRule[Mbr|M\rightrule]%
{$\ddiamond{}{}/\dbox{}{}$ generalization=M=G+K} 
{\linferenceRule[sequent]
  {\lsequent[L]{} {\dbox{\ausprg}{\busfml}} 
  &\lsequent[g]{\busfml} {\ausfml}}
  {\lsequent[L]{} {\dbox{\ausprg}{\ausfml}}}
}{}%
\]
The remaining four branches of arithmetic and the differential equation property
\(\lsequent{\invhv{x}{v}} {\dbox{\odehvdef{x}{v}}{\invhv{x}{v}}}\)
make it fairly easy to prove or disprove concrete invariant candidate choices for $\invhv{x}{v}$ using decision procedures for real arithmetic \cite{DBLP:journals/jsc/CollinsH91} and either solution axioms or \dL proofs for ODE invariants \cite{DBLP:conf/lics/PlatzerT18}.
It is also not difficult to generate a suitable invariant $\invhv{x}{v}$ from the desired ODE invariance property \cite{Platzer18}:
\begin{equation*}
\invhv{x}{v} \mdefequiv 
\invhvdef{x}{v}
\end{equation*}
This invariant makes it possible to complete the bouncing ball proof with the constant assumptions \(c=1 \land g>0\), which are easily added \cite{Platzer18}.
\end{example}

Working with the solution \(x(t)=x+vt-\frac{g}{2}t^2, v(t)=v-gt\) suffices for proving all differential equation properties arising in the above example.
But differential invariants enable a significantly easier proof from the differential equation itself instead of its more complicated solution.
The following special case of the more general differential invariants proof rule suffices here:
\[
\cinferenceRule[dIeq|dI]{differential invariant}%
{\linferenceRule[sequent]
  {\lsequent[g]{\ivr}{\Dusubst{\D{x}}{\genDE{x}}{\der{\astrm}=\der{\bstrm}}}}
  {\lsequent{\astrm=\bstrm}{\dbox{\pevolvein{\D{x}=\genDE{x}}{\ivr}}{\astrm=\bstrm}}}
}{}
\]
Here, \irref{dIeq} shows that an equality \(\astrm=\bstrm\) always remains true after an ODE \(\D{x}=\genDE{x}\) by proving that its differentials $\der{\astrm}$ and $\der{\bstrm}$ are equal after assigning $\genDE{x}$ to $\D{x}$. 
\begin{sequentdeduction}[array]
\linfer[dIeq]
{\linfer[Dassignb]
  {\linfer[qear]
    {\lclose}
    {\lsequent{x\geq0} {2gv=-2v(-g)}}
  }
  {\lsequent{x\geq0} {\Dusubst{\D{x}}{v}{\Dusubst{\D{v}}{-g}{2g\D{x}=-2v\D{v}}}}}
}
{\lsequent{2gx=2gH-v^2} {\dbox{\pevolvein{\D{x}=v\syssep\D{v}=-g}{x\geq0}}{2gx=2gH-v^2}}}
\end{sequentdeduction}

\subsection{Soundness and Completeness}

While not the focus of this overview, differential dynamic logic also enjoys very strong theoretical properties.
The most important property of soundness (all formulas that have a \dL proof are valid) is a \emph{conditio sine qua non}, a condition without which logic could not be. And, of course, soundness holds for the \dL calculus \cite{DBLP:journals/jar/Platzer08,DBLP:conf/lics/Platzer12b,DBLP:journals/jar/Platzer17} as has even been cross-verified in both Isabelle/HOL and Coq \cite{DBLP:conf/cpp/BohrerRVVP17}.
The converse question of completeness, i.e., whether all true \dL formulas have a \dL proof, is more complicated.
It is answered in the affirmative both for completeness relative to differential equations and relative to discrete dynamics (precise statements can be found in the literature \cite{DBLP:journals/jar/Platzer08,DBLP:conf/lics/Platzer12b,DBLP:journals/jar/Platzer17}).

\begin{theorem}[Continuous relative completeness of \dL \cite{DBLP:journals/jar/Platzer08,DBLP:conf/lics/Platzer12b}] \label{thm:dL-complete}
  \index{complete!relatively!dL@\dL}
  The \dL calculus is a \emph{sound and complete axiomatization} of hybrid systems relative to differential equations, i.e.,
  every valid \dL formula can be derived in the \dL calculus from valid differential equation tautologies.
\end{theorem}

\begin{theorem}[Discrete relative completeness of \dL \cite{DBLP:conf/lics/Platzer12b}] \label{thm:dL-complete2}
  \index{complete!relatively!dL@\dL}
  The \dL calculus is a \emph{sound and complete axiomatization} of hybrid systems relative to discrete dynamics, i.e.,
  every valid \dL formula can be derived in the \dL calculus from valid discrete system tautologies.
\end{theorem}

In particular, relative completeness theorems (which have constructive proofs) imply that the required (variants or) invariants are always expressible in \dL to succeed with the proofs.
If the invariants are in real arithmetic, the remaining questions about differential equations are completely provable in the \dL calculus (again referring to the literature \cite{DBLP:conf/lics/PlatzerT18} for precise statements):

\begin{theorem}[Invariant completeness of \dL \cite{DBLP:conf/lics/PlatzerT18}] \label{thm:dL-invariant-complete}
  The \dL calculus is a \emph{sound and complete axiomatization} of arithmetic invariants of differential equations, i.e.,
  all true real arithmetic invariants of differential equations are provable in \dL (and all false ones are disprovable in \dL).
\end{theorem}

Moreover, real arithmetic invariants of differential equations are decidable directly by a derived axiom of \dL \cite{DBLP:conf/lics/PlatzerT18}.
In particular, in the common case where the respective hybrid system invariants that always exist in \dL by \rref{thm:dL-complete} for valid safety formulas are actually real arithmetical, \rref{thm:dL-invariant-complete} guarantees that the resulting differential equation properties are provable in \dL if and only if they are true.
In theory, the only thing that can go wrong in a hybrid systems safety proof of a valid formula is that the required invariants exist in \dL but are not real arithmetical.
In practice, what can still go wrong, in addition, is that scalability limits make the resulting computations infeasible.
This is where user guidance in a theorem prover can help get proofs unstuck with user insights about the CPS design that the proof automation fails to find.

\subsection{\dL Proofs in The \KeYmaeraX Theorem Prover}

The differential dynamic logic proof calculus is implemented in the theorem prover \KeYmaeraX \cite{DBLP:conf/cade/FultonMQVP15} (along with an implementation of differential game logic for hybrid games \cite{DBLP:journals/tocl/Platzer15} that is not the focus here).
The prover implements the uniform substitution calculus for differential dynamic logic \cite{DBLP:journals/jar/Platzer17} (and differential game logic \cite{DBLP:conf/cade/Platzer18,DBLP:conf/cade/Platzer19}).
The advantage of a uniform substitution design is that this leads to a minimal soundness-critical core consisting only of a verbatim copy of each axiom formula besides the uniform substitution application mechanism and renaming.
For performance reasons, \KeYmaeraX also implements a propositional sequent calculus with Skolemization \cite{DBLP:journals/jar/Platzer08} to efficiently normalize logical formulas.
This gives \KeYmaeraX a minimal soundness-critical core of under 2000 lines of code.

At the same time, \KeYmaeraX provides a language for custom tactics \cite{DBLP:conf/itp/FultonMBP17}, sophisticated proof automation, including invariant generation \cite{DBLP:conf/fm/SogokonMTCP19}, and a versatile user interface \cite{DBLP:conf/fide/MitschP16}, but those are markedly outside the soundness-critical part of \KeYmaeraX even if still just as important for practical verification in \KeYmaeraX.

For deciding formulas of real arithmetic, \KeYmaeraX calls Mathematica \cite{Mathematica12} or Z3 \cite{DBLP:conf/tacas/MouraB08,DBLP:conf/cade/JovanovicM12}.
Instead of trusting those solvers, polynomial witnesses can be used to rigorously prove the validity of universal real arithmetic formulas \cite{DBLP:conf/tphol/Harrison07,DBLP:conf/cade/PlatzerQR09}.
In theory, that is complete for universal real arithmetic but only works in practice for medium complexity problems.
While proof-producing real arithmetic decision procedures exist \cite{DBLP:conf/cade/McLaughlinH05}, they are only used in the old \KeYmaera 3 prover \cite{DBLP:conf/cade/PlatzerQ08}, because their scalability is limited to low-dimensional problems.
In the older \KeYmaera 3, real arithmetic can also be decided with virtual substitution \cite{DBLP:journals/aaecc/Weispfenning97} using Redlog \cite{DBLP:journals/sigsam/DolzmannS97}, which works in cases where the polynomial degrees remain at most cubic.
Generally, however, more scalable verification of nonlinear real arithmetic remains an important future challenge.

\section{ModelPlex: Model Safety Transfer}

\KeYmaeraX provides automatic and interactive and programmable ways for rigorously proving the safety (or other properties) of hybrid systems in differential dynamic logic.
However useful and reassuring that is for the hybrid system and its control, it is important to realize that the outcome is still a statement about the hybrid system, which is a mathematical model of the cyber-physical system.
A safety proof for a hybrid system only implies the safety of a cyber-physical system to the extent that the hybrid system is an adequate model of the CPS.
If we rename a \KeYmaeraX input file from \texttt{bouncing-ball.kyx} to \texttt{car.kyx} and prove it, then we have not made any actual car any safer, because bouncing ball models are quite inappropriate models for cars.
In a sense, of course, the appropriate course of action is that we then need to get a better model and put it into \texttt{car.kyx}.
But how do we then really argue that a safety proof of the model in the improved \texttt{car.kyx} file entails that the car is safe while a safety proof of the \texttt{bouncing-ball.kyx} model did not?
Or what would it take to argue that \texttt{bouncing-ball.kyx}  at least is a good model for bouncing balls (spoiler alert: it is not as long as we keep the assumption $c=1$).
This requires us to prove that we \emph{built the right model} (does it suitably model all real behavior) and not just that the \emph{model was built right} (its controllers work safely in that model).

Proving safety of a real CPS is surprisingly subtle, precisely because it does not help to fall back to even more models along the way that need to be justified, too, to accurately reflect reality.
The solution to cut through this Gordian knot of model mysteries is \emph{ModelPlex}, which provides a way to rigorously transfer safety proofs of hybrid systems models to safety results about real CPS implementations \cite{DBLP:journals/fmsd/MitschP16}.
The way it works is that ModelPlex conducts a proof in differential dynamic logic about the safe behavior of the real CPS that combines the \dL safety proof about the hybrid systems model with proofs obtained online from monitors that witness the compatibility of the observed real executions of the CPS with the safety-critical part of the hybrid systems model.
This results in a proof in which the majority of the work is performed offline ahead of time, while a few critical measurements performed by monitors fill in online at runtime the parts of the proof that link to reality and are unprovable offline.
But the required monitors are proved safe in \dL ahead of time and synthesized along with their correctness proof in \KeYmaeraX from the safety proof of the original \dL formula.

\begin{figure}[htb]
  \centering
  \includegraphics[height=2.7cm]{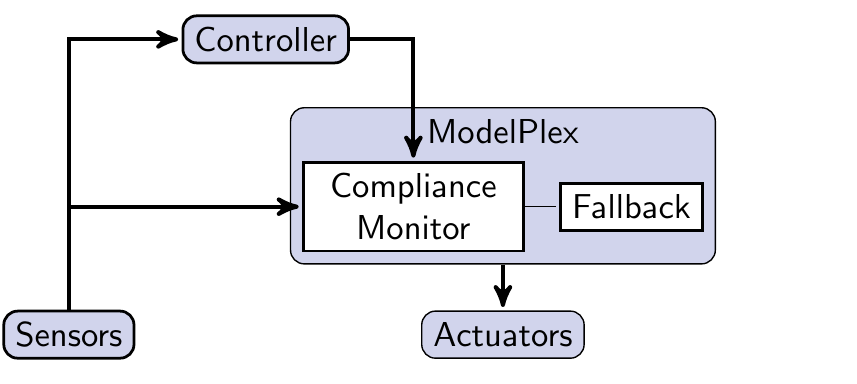}
  \caption{ModelPlex monitor sits between controller and actuator to check the controller's decisions for compliance with the model based on sensor data with veto leading to a safe fallback}
  \label{fig:ModelPlex-logo}
\end{figure}

When run on the CPS, synthesized ModelPlex monitors check model compliance and safeguard decisions obtained by the unverified controller implementation (\rref{fig:ModelPlex-logo}).
In that respect, ModelPlex monitors are used like Simplex monitors \cite{DBLP:conf/acc/SetoKSC98}, except that ModelPlex monitors are proved correct, synthesized from the verified models, and not only worry about the safety of the controls but also ensure safe compliance with the whole model (hence the name \emph{Model}Plex).

\begin{example}[Correct controller monitor] \label{ex:bouncing-ball-controller-monitor}
A provably correct runtime monitor for the bouncing ball \emph{controller} from \rref{ex:bouncing-ball-dL} is as follows \cite[Ex.\,19.3]{Platzer18}:
\begin{equation*}
\big(x=0\land\posterior{v}=-v ~\lor~ x>0\land\posterior{v}=v\big) \,\land\, \posterior{x}=x
\end{equation*}
This formula expresses a relation for acceptable transitions of the controller implementation from old values of $x$ and $v$ to new values $\posterior{x}$ and $\posterior{v}$.
\end{example}

\begin{example}[Correct model monitor] \label{ex:bouncing-ball-model-monitor}
A provably correct runtime monitor for the bouncing ball \emph{model} from \rref{ex:bouncing-ball-dL} is the following \cite[Prop.\,19.1]{Platzer18}:
\begin{equation*}
\begin{aligned}
&2g(\posterior{x}-x)=v^2-(\posterior{v})^2\land x\geq0 \land (\posterior{x}>0 \land \posterior{v}\leq v \lor \posterior{x}=0 \land \posterior{v}\geq-v)
\end{aligned}
\end{equation*}
\end{example}

Monitors get even more subtle with the uncertainty of partially observable hybrid systems, but ModelPlex generalizes suitably \cite{DBLP:journals/corr/abs-1811-06502}.
The point is, however, that both monitors from Examples~\ref{ex:bouncing-ball-controller-monitor} and~\ref{ex:bouncing-ball-model-monitor} are violated when comparing the behavior with a real bouncing ball, whether uncertainty is taken into account or not.
This shows, for example, that $c=1$ is an unrealistic assumption for a true damping coefficient, which, after all, makes a real bouncing ball bounce back up less over time (recall \rref{fig:bouncingball-simple-trajectory}).

\section{VeriPhy: Executable Proof Transfer}

The ModelPlex monitors synthesized for safeguarding the transfer of safety properties of a hybrid system to a real CPS are a crucial ingredient.
But they are also not directly executable without error on a real CPS.
The reason is that ModelPlex monitors are correct mathematical conditions for the safety transfer from models to reality, but are still phrased in real arithmetic.
Real arithmetic provides the right setting in which to establish the safety of motion in the real world from one real-valued position to another connected by the real-valued differential equations of the model.
But such infinite precision real arithmetic is hardly executable on any actual CPU in any reasonable amount of time (or at all).
There is also still a gap between knowing the right formulas to monitor from ModelPlex and being able to run proper safe executables on the computation devices of a CPS.

This is where \emph{VeriPhy} comes in, the verified pipeline with which verified controller executables can be generated from verified cyber-physical system models \cite{DBLP:conf/pldi/BohrerTMMP18}.
VeriPhy adds machine executability in provably correct ways to the mathematical accuracy of ModelPlex monitors.
It turns a hybrid systems model verified safe in \KeYmaeraX into executable machine code for a sandbox that inherits safety for the real CPS in provably safe ways with a chain of proofs in theorem provers.
The resulting CPS sandbox checks and corrects all actions of an unverified CPS controller provided separately, and safely flags plant violations when possibly unsafe deviations are detected compared to the safe behavior of the model.
The last step of the VeriPhy pipeline compiles provably correctly to CakeML \cite{DBLP:conf/popl/KumarMNO14}, thanks to which machine code can be generated provably correctly for multiple CPUs, e.g., x64 and arm6 architectures.

At this point, one might take the bouncing ball model from \rref{ex:bouncing-ball-dL} that was verified in \rref{ex:bouncing-ball-dL-proof} and generate provably correct machine code for it with VeriPhy, which will use the synthesized ModelPlex controller and plant monitors from Examples~\ref{ex:bouncing-ball-controller-monitor} and~\ref{ex:bouncing-ball-model-monitor}, respectively.
But, admittedly, purchasing a genuine bouncing ball would have led to an easier and more realistic execution platform, which, furthermore, would not have fallen prey to the deficiencies caused immediately by the unrealistic assumption $c=1$.
Yet, that is a consequence of the fact that the discrete operations in \rref{ex:bouncing-ball-dL} were just a description of bouncing ball physics, and not within the purview of an ingenious cyber controller that is in active need to control how balls bounce.
The discrete control of a robot's ping-pong paddle to react upon the motion of a ping-pong ball \cite{Platzer18} would have led to a related model in which VeriPhy would serve a better purpose.

\section{Safe Learning in CPS}

The logical foundations that have been surveyed thus far give a solid basis for the development of provably correct cyber-physical systems \cite{Platzer18}.
But an additional twist arises for \emph{autonomous cyber-physical systems} \cite{DBLP:conf/qest/Platzer19} that achieve smart and largely unsupervised decisions on their own using machine learning and artificial intelligence (AI) techniques.
The extra complication is that their behavior is not just a consequence of their programming, but also of their experience and the data they learned from, which is stored, e.g., in policy tables that influence their future decisions \cite{Russel:95:AIAMA}.
Lauded for their flexibility, this AI phenomenon, however, makes the behavior of AI algorithms much harder to predict, which is adverse to their safety.

Technically, one could embed all operations that the AI performs explicitly into the discrete part of a hybrid program.
In practice, however, a rendition of the safety-critical data and code then quickly reaches completely infeasible sizes.
Due to its use of Markov Decision Process policy optimization, an explicit rendition of the Next-generation Airborne Collision Avoidance System ACAS~X \cite{KochenderferHC12}, for example, would have led to a hybrid program with about half a trillion cases.
That is why ACAS~X was verified in \KeYmaeraX by splitting off the policy comparison from a general symbolic verification \cite{DBLP:journals/sttt/JeanninGKSGMP17}.

More generally, AI and machine learning can be understood as unverified implementations whose compliance with provably safe operations of the CPS can be ensured using ModelPlex \cite{DBLP:journals/fmsd/MitschP16}.
While its primary intent is to provide ways of bridging safety of models to safety of implementations, ModelPlex does not actually mind for what reasons part of the CPS are unverified, including for complexity reasons or due to the use of AI and learning in the CPS.

\paragraph{Sandboxed Execution of a Learned CPS.}
Combining reinforcement learning with ModelPlex safety monitors enables provably safe reinforcement learning in CPS \cite{DBLP:conf/aaai/FultonP18}.
Reinforcement learning repeatedly chooses actions and observes outcomes in the (simulated or real) environment and makes those actions more likely if the outcome was favorable and less likely if it was not \cite{sutton-barto:1998a}.
The most obvious way of benefiting from a provably safe ModelPlex monitor in a learning CPS is to leave learning alone while training and then, during deployment, simply safeguard all actions performed by the trained CPS using the ModelPlex monitors.
The advantage of this approach is that it is conceptually simple, but the downside is that the experience obtained from the learning algorithm will be used in a real environment that differs from the training environment, precisely because the ModelPlex monitors only interfere after deployment not during training, which may lead to a suboptimal performance.

\paragraph{Sandboxed Learning of a CPS.}
Another alternative is to integrate reinforcement learning with ModelPlex safety monitors more deeply and use the feedback from safety monitors throughout the learning \cite{DBLP:conf/aaai/FultonP18}.
First of all will the reinforcement learning algorithm learn to act in the same sandboxed environment that it would otherwise be acting in after deployment.
Second of all does that give the learning algorithm a chance to learn the concept of safety and learn which actions will be vetoed due to ModelPlex monitor violations.
Furthermore can the reinforcement learning algorithm benefit from the early feedback that the ModelPlex monitors provide, which immediately flag actions as problematic if they could have unsafe downstream consequences, instead of having to wait for a scenario where an actual disaster happened later on, and then find out how to propagate this outcome back to the earlier root cause decision.
Such early feedback has been observed to significantly speed up convergence of reinforcement learning.
Another advantage of sandboxed learning is that the resulting policy (when started from a safe initial policy or fallback) can be provably safe \cite{DBLP:conf/aaai/FultonP18}.
A downside of sandboxed learning is that the learning may become overly reliant on the sandbox interference, and that it cannot learn behavior outside the known safe sandbox.
This can be compensated for by allowing exploration outside known safe actions (within a \emph{simulated} training environment).

\paragraph{Recovery Learning of a CPS.}
Yet another possibility for combining formal methods with machine learning is to continue to use information from safety proofs even outside known safe parts of the world \cite{DBLP:conf/aaai/FultonP18}.
In that case do quantitative versions of ModelPlex monitors serve as reward signals that tend to pull the system back from unsafe into safe space.
Beyond the experimental observation that this enables a safe recovery outside well-modeled parts of the world is it possible to give rigorous safety proofs of the resulting behavior of the learning CPS for the case of multiple possible models that are not all wrong or that can be modified to safely fit reality with \emph{verification-preserving model updates} \cite{DBLP:conf/tacas/FultonP19}.
The basic idea is to use the conjunction of all ModelPlex monitors of plausible models to determine which action is safe while discarding models whose predictions did not end up happening.
Satisfying assignments to logical combinations of ModelPlex monitors of different models can be used to actively plan differentiating experiments and converge a.s.\ to the true model.

Of course, all three of these types of modifications of learning algorithms or deployments of learned systems could be used with any other monitor, but then the crucial ingredient of provable safety of the outcome would be missing.

There is no point in using any learning to find out how to act in the bouncing ball model from \rref{ex:bouncing-ball-dL}, because its discrete part is a deterministic if-then-else statement allowing no room for any actual decisions.
But suppose the ground is replaced by a ping-pong paddle that can actively hit the ball harder than a passive floor could.
Then, there suddenly is a control choice to be made and it can make sense to use learning from experience to find out how hard to best hit the ball under what circumstance.
This would require a change in the model replacing the fixed damping coefficient $c$ by a variable pushback from the ground as a function of how quickly the ping-pong paddle swings.

\section{Conclusion}

While short and succinct out of necessity, the purpose of this article was to survey the Logical Foundations of Cyber-Physical Systems approach, which is detailed in a recent textbook of the same name \cite{Platzer18}.
This approach provides a solid foundation for the analysis and correct development of cyber-physical systems and their quite subtle controllers.
The \KeYmaeraX theorem prover provides automatic and interactive and programmable ways for rigorously proving the safety (or other properties) of hybrid systems in differential dynamic logic.
Extensions were summarized that transfer safe models to a safe reality, provide provably safe execution on CPS, or augment safe CPS with safe learning to obtain the best of both worlds.
Extensions from hybrid systems to hybrid games are also implemented in the theorem prover \KeYmaeraX, but presented along with their generalizations elsewhere \cite{DBLP:journals/tocl/Platzer15,DBLP:journals/tocl/Platzer17,Platzer18}.
Extensions from hybrid systems to distributed hybrid systems \cite{DBLP:journals/lmcs/Platzer12b} are fascinating and useful, but presently only implemented in previous-generation theorem provers \cite{DBLP:conf/cade/PlatzerQ08,DBLP:conf/icfem/RenshawLP11}.

\noindent\textbf{Acknowledgements.}
I thank Brandon Bohrer for his feedback on this overview.

\bibliographystyle{apalike}
\bibliography{platzer,bibliography}
\end{document}